\documentclass{iopart}
\def\be{\begin{equation}}
\def\ee{\end{equation}}
\def\bea{\begin{eqnarray}}
\def\eea{\end{eqnarray}}
\usepackage{iopams}
\usepackage[dvips]{color}
\def\beq{\begin{equation}}
\def\eeq{\end{equation}}
\def\bea{\begin{eqnarray}}
\def\eea{\end{eqnarray}}

\begin{document}
\title[Generalized BTZ]
{Functional determinants, generalized BTZ geometries and Selberg zeta function}

\author{R Aros and D E D\'{\i}az} 

\address{Universidad Andres Bello,
Departamento de Ciencias Fisicas,
Republica 220, Santiago, Chile}
\ead{raros,danilodiaz@unab.cl}

\begin{abstract}

We continue the study of a special entry in the AdS/CFT dictionary, namely a
'holographic formula' relating the functional determinant of the scattering
operator in an asymptotically locally anti-de Sitter
(ALAdS) space to a relative functional determinant of the scalar Laplacian in the bulk.
A heuristic derivation of the formula involves a one-loop quantum effect in the bulk and the corresponding sub-leading correction at large N on the boundary.

We presently explore a higher-dimensional version of the spinning BTZ black hole obtained as a quotient of hyperbolic space by a discrete subgroup of isometries generated by a loxodromic (or hyperbolic) element consisting of dilation (temperature) and torsion angles (spinning).
The bulk computation is done using heat-kernel techniques and fractional calculus. At the boundary, we get a recursive scheme that allows us to range from the non-spinning to the full-fledged spinning geometries. The determinants are compactly expressed in terms of an associated (Patterson-)Selberg zeta function and a connection to quasi-normal frequencies is discussed.

\end{abstract}


\section{Introduction}

The \textit{AdS/CFT correspondence} has been a very productive area of research ever since its appearance, more than a decade ago, in the form of Maldacena's conjecture together with a calculational prescription~ \cite{Maldacena:1998re,Gubser:1998bc,Witten:1998qj}. Many developments are, perhaps more appropriately, embraced under the name \textit{gauge/gravity duality} for they depart from the original canonical examples that involved anti-de Sitter spacetime and the bulk side is usually restricted to the (super)gravity approximation. On one hand, these canonical (or most symmetric) cases seem to be grounded in solid mathematical foundations, such as harmonic analysis on symmetric spaces~\cite{Camporesi:1990wm}, representation theory~\cite{Dobrev:1998md}, hyperbolic and conformal geometry~\cite{fefferman-2007}.
On the other hand, recent and exciting applications to physically relevant situations (see e.g.~\cite{McGreevy:2009xe,Horowitz:2006ct}) are more heuristic, and therefore mathematically exact results become rare. This starts already when including finite temperature on the boundary theory, leading to AdS black holes as bulk background geometry.

One notable exception is the BTZ black hole~\cite{Banados:1992wn}. While having essentially the same standard features of higher-dimensional AdS black holes, its virtue of being a space of constant negative curvature greatly simplifies many derivations (cf. ~\cite{Carlip:1995qv}). Here one can include temperature and still obtain explicit analytic results for the Green's functions, spectrum of quasi-normal modes, etc. and several holographic features have a mathematical counterpart~\cite{Birmingham:2001dt,Manin:2002hn}.

Another instance that transcends its original frame is the case of a holographic formula relating the functional determinant of the scattering operator in an asymptotically locally anti-de Sitter(ALAdS) space to a relative functional determinant of the scalar Laplacian in the bulk.
The conformal anomaly at the boundary at leading large N can be read off from the regularization of the classical gravitational action; a quantum correction to this classical gravitational action such as the one-loop contribution of a scalar field corresponds to a sub-leading effect at the boundary. This was confirmed in \cite{Gubser:2002zh,Gubser:2002vv}. However, the mapping can be further extended to an equality between functional determinants~\cite{Hartman:2006dy,Diaz:2007an}, namely
\begin{equation}\label{hol-for}
\frac{\det_{-}(\Delta_X-\lambda(n-\lambda))}{\det_{+}(\Delta_X-\lambda(n-\lambda))}
=\det\,S_M(\lambda),
\end{equation}
where $\Delta_X$ is the Laplacian operator in the bulk and $S_M(\lambda)$ stands for the two-point correlation function of the boundary operator $O_{\lambda}$ dual to the bulk scalar field $\Phi$.
The label $+$ refers to the usual determinant, obtained for example via heat kernel techniques, whereas the label $-$ refers to the analytic continuation from $\lambda=\lambda_+$ to $\lambda_-=n-\lambda$.

In this work, we bring together these two instances to explore the holographic formula in a generalized BTZ geometry. This has been previously done for a non-spinning BTZ geometry with spherical horizon~\cite{diaz-2009-42}, but it was realized that the bulk computation can be readily extended to the more general case covered by a result due to Patterson~\cite{PattersonKleian89} in terms of an associated Selberg zeta function for the resulting spinning BTZ geometry. The challenge we presently undertake is to elucidate how this general result can be recovered on the boundary.

Regarding the `separate lives' of each side of the holographic formula, a few remarks are worth mentioning. The functional determinant on the bulk side is the central object when computing one-loop effective actions~\footnote{As said, one-loop corrections correspond to sub-leading terms in the large-N expansion on the boundary and a any refinement of the holographic correspondence to include them leads to such bulk determinants. This opens the possibility to find interesting effects in strongly coupled field theories, not captured by the leading order, as in~\cite{Denef:2009yy}.}. For example, in BTZ$_{3}$ the effective action has been computed by Mann and Solodukhin~\cite{Mann:1996ze}, and later Perry and Williams noticed the connection with Selberg zeta function.
The functional determinant on the boundary, in turn, involves a pseudo-differential operator and  has been less explored in the physical literature; however, it is related to conformal powers of the Laplacian (cf.~\cite{Diaz:2007an}) and one can find a connection with Selberg zeta function in multi-loop amplitudes for the bosonic string~\cite{D'Hoker:1985pj}. AdS/CFT correspondence, via the holographic formula in $BTZ_3$, connects these two a priori unrelated computations.

The outline of this paper is as follows.  The quotient geometry and a general ansatz for the bulk metric is reviewed in the next section. The computation of the determinant on the RHS of the holographic formula, using a particular parametrization of the compact boundary, is performed in section 3. For completeness, a new derivation of the result by Patterson for the LHS, using this time heat-kernel and fractional calculus, is included in section 4. Section 5 elaborates on the representations of the Selberg zeta function that encodes both sides of the formula in the present case, and elucidates the connection with quasi-normal frequencies. Section 6 includes concluding remarks and perspectives. Three appendices provide some supplementary material.

\section{Quotient geometry}

The existence of the BTZ solution in the form of a quotient in three dimensions streamed out the search for higher dimensional analogs \cite{Aminneborg:1996iz,Aminneborg:1997pz,Banados:1998dc,Krasnov:2000zq}. Indeed, one can foresee, by the rich geometrical structure of the negative curvature manifolds such as AdS$_{d}$ \cite{BookHyperbolicManifolds}, the presence of interesting solutions of the form $\Gamma \backslash  \textrm{AdS}_{d}^*$, where AdS$^*_d$ is a section of AdS$_d$ and $\Gamma$ a subgroup of the AdS isometry group. A classification of the possible identifications can be found in \cite{FigueroaOFarrill:2004bz,FigueroaOFarrill:2004yd} and references therein. Locally, these spaces can be written as
\begin{equation}\label{GeneralSpace}
ds^{2}=B(r)^{2}(d\hat{x}^{i}d\hat{x}^{j}\hat{g} _{ij})+ C(r)^{2}dr^{2}+A(r)^{2}(%
d\tilde{y}^{m} d\tilde{y}^{n} \tilde{g}_{mn})
\end{equation}
where $\hat{g}_{ij}$ and $\tilde{g}_{mn}$ are intrinsic metric of two constant curvature manifolds, hereinafter referred to as worldsheet and transverse section, respectively.

Among the different possible cases, in this work will be considered only a subclass. In terms of the Poincar\'e half plane,
\begin{equation}\label{Poincare}
    ds^2=\frac{dz^2+d\overrightarrow{x}^2}{z^2},
\end{equation}
the identification $\Gamma$ is generated by a hyperbolic element of the discrete group of isometries whose action is
\begin{equation}\label{Identification}
(z,\overrightarrow{x})\sim e^l(z,\mathbb{A}\overrightarrow{x}).
\end{equation}
Here $\mathbb{A}$ is a rotation, which can be casted in block form with eigenvalues $e^{\pm i \varphi_k}$ with $k=1,...,K\leq\lfloor n/2\rfloor$ \footnote{$\lfloor \alpha \rfloor$ stands for the largest integer smaller than or equal to $\alpha$} and an extra unit eigenvalue in case $n$ is odd. In geometric terms, to cast $\mathbb{A}$ in block form is equivalent to reduce the rotations to planes that do not intersect, thus $k$ also labels the different planes where the rotations occur. If $\mathbb{A}$ is trivial then the transverse section in Eq.(\ref{GeneralSpace}) corresponds to a sphere. Otherwise the transverse section and
worldsheet are intertwined manifolds.

In order to classify this identification \cite{FigueroaOFarrill:2004bz} one can notice that Eq.(\ref{Identification}) is generated by the Killing vector
\begin{equation}\label{KillingVector}
    \xi_{BTZ} = l \left(z\frac{\partial}{\partial z} + x^{i} \frac{\partial}{\partial x^{i}}\right)  + \sum_{k=1}^{\lfloor n/2\rfloor} \varphi_{k} \left(x^{k_{1}}\frac{\partial}{\partial x^{k_{2}}} - x^{k_{2}}\frac{\partial}{\partial x^{k_{1}}}\right).
\end{equation}
where the last part of the vector stands for the sum of the generators of a rotation in the each of the $(x^{k_{1}},x^{k_{2}})$ planes.

There are several known examples. For instance one can quote \cite{Banados:1997df} whose line element is given by
\begin{equation}\label{ds/sch}
ds^2 = N(r)^2 d\Sigma_3 +  N(r)^{-2} dr^2 + r^2 d\phi^2,
\end{equation}
with $N^2(r) = r^2 -r_+^2$ and
\begin{equation}\label{sph}
d\Sigma_3 = - \cos^2{\theta}\, dt^2 + \frac{1}{r_+^2} (d\theta^2 +
\sin^2{\theta} d\chi^2).
\end{equation}
The horizon in these coordinates is located at $r=r_+$. In this case the identification is performed along the Killing vector (\ref{SimplestKilling})
\be
\xi = 2\pi r_{+} \frac{\partial}{\partial \phi} = 2\pi r_{+} \left( w\frac{\partial }{\partial p} + p\frac{\partial }{\partial w} \right)= 2\pi r_{+}  \left( z\frac{\partial }{\partial z} + y^i \frac{\partial }{\partial y^i} \right),
\ee
where the coordinates $w,p$ are defined in \ref{Coordinates}. In terms of the general identification (Eq.(\ref{Identification})) this corresponds to the case $\mathbb{A}$ trivial.

\section{Boundary}

Let us start with the functional determinant on the boundary. We first need to compute the two-point correlation function of the dual operator with scaling dimension $\lambda$. A useful way to get this information is to examine the scattering of the bulk field in the black hole background.

\subsection{Scattering}

The metric is first suitable written in ``scattering form''
\be
g_{X}=dt^2+ \mbox{ch}^2t\,du^2+ \mbox{sh}^2t\,d\Omega^2_{n-1}~,
\ee
so that the (positive) Laplacian takes the following form
\be
\Delta_X=-\frac{1}{\mbox{ch}t\,\mbox{sh}^{n-1}t}\,\partial_t\left(\mbox{ch}t\,\mbox{sh}^{n-1}t\,\partial_t\right)
-\frac{1}{\mbox{ch}^2t}\partial_u^2-\frac{1}{\mbox{sh}^2t}\Delta_{\Omega}~.
\ee
One has to consider eigenfunctions of the form $e^{i\kappa u}\times$\{\textit{angular part}\}, compatible with the identifications. For instance, when there is no mixing $\kappa=2\pi N/l,\, N\in \mathbb{Z}$; with only one block, $\kappa=2\pi N/l\pm r\varphi/l$ and so on.
This leads then to an effective one-dimensional operator
\be\fl
\widehat{H}_{L,\kappa}=-\frac{1}{\mbox{ch}t\,\mbox{sh}^{n-1}t}\,d_t\left(\mbox{ch}t\,\mbox{sh}^{n-1}t\,d_t\right)
+\kappa^2\,\mbox{sech}^2t+\gamma^2_L\,\mbox{csch}^2t~,
\ee
where $\gamma^2_L=L(L+n-2)$ are the eigenvalues of the angular part.
Here one can recognize a one-dimensional stationary Schr\"odinger equation with a (generalized) P\"oschl-Teller potential
\be
H_{L,\kappa}=-d^2_t+\alpha(\alpha+1)\,\mbox{csch}^2t-\beta(\beta+1)\,\mbox{sech}^2t
\ee
with $\alpha=-3/2+n/2+L$ and $\beta=-1/2+i \mid\kappa\mid$.

The scattering matrix~\cite{PerryandWillians} (here $\lambda=n/2+\nu$) is known to be given, modulo unimportant factors, by
\be\fl\label{scattering}
\mathbb{S}_{L,\kappa}(\lambda)=\frac{\Gamma(L/2+n/4+\nu/2+i\kappa/2)\;\Gamma(L/2+n/4+\nu/2-i\kappa/2)}
{\Gamma(L/2+n/4-\nu/2+i\kappa/2)\;\Gamma(L/2+n/4-\nu/2-i\kappa/2)}~.
\ee
These are the coefficients of the expansion of the scattering operator, or thermal two-point correlation function, in the chosen basis.

\subsection{Sphere: eigenfunctions}

As a technical prelude, however, one has to learn now how to construct eigenfunctions of the Laplacian on the $(p+q+1)$-sphere out of the usual harmonics for the $p-$ and $q-$sphere. This is very similar to what we had before for the bulk metric in scattering form,
\be ds^2=d\theta^2+\mbox{cos}^2\theta d\Omega^2_p+\mbox{sin}^2\theta d\Omega^2_q~,\ee
with the Laplacian
\be\fl
\Delta_{\Omega_{p+q+1}}=\frac{1}{\mbox{cos}^{p}\theta\,\mbox{sin}^{q}\theta}\,\partial_{\theta}\left(\mbox{cos}^{p}\theta\,\mbox{sin}^{q}\theta
\,\partial_{\theta}\right)+\frac{1}{\mbox{cos}^2\theta}\Delta_{\Omega_p}+\frac{1}{\mbox{sin}^2\theta}\Delta_{\Omega_q}~.
\ee
Plugging in the eigenvalues of the spherical harmonics for the little spheres (orbital quantum numbers $r$ and $s$) we end up with an effective one-dimensional equation which can be eventually related to Jacobi polynomials of order  $m=0,1,2,...$. The relation with the orbital quantum number $L$ of the $(p+q+1)-$sphere is $L=2m+r+s$.
Here one can check that the counting of states and degeneracies agree.

\subsection{Spinning BTZ: one block}

The functional determinant we need to compute on the boundary can be casted as a trace

\be
\mbox{log\,det\;}\mathbb{S}= \mbox{tr\,log\;}\mathbb{S}=\sum_{eigenstates}\mbox{log\;}\mathbb{S}_{L,\kappa}~.
\ee
Now consider one rotation-block. Following essentially the same steps as in the non-spinning case~\cite{diaz-2009-42}, i.e. taking derivative and using an integral representation of the gamma function, we write for the derivative of log-det of the scattering operator

\be\fl
\mbox{tr}(\mathbb{S}^{-1}\partial_{\lambda}\mathbb{S}(\lambda))=-\sum_{eigenstates}\;
\int^{\infty}_0 dt \frac{e^{-tL/2-it\kappa/2}}{1-e^{-t}}(e^{-\lambda t/2}+e^{(\lambda-n)t/2})
\ee
but now $\kappa=2\pi N/l\pm r\varphi/l$ depends not only on $N$ but on the orbital quantum number on the $S^1$ and $S^{n-3}$. The $S^{n-1}$ is decomposed in an $S^1$, an $S^{n-3}$ and a polar angle $\theta$, so that $L=2m+r+s$.
The trick is to sum over $m,r,s=0,1,2,3,...$ taking into account the degeneracies
\bea\label{eigen}
\sum^{\infty}_{N=-\infty}e^{-i\pi Nt/l}
\sum^{\infty}_{m=0}e^{-tm}\times\nonumber\\
\left(1+\sum^{\infty}_{r=1}e^{-rt/2}[e^{ir\varphi t/2l}+e^{-ir\varphi t/2l}]\right)
\;\sum^{\infty}_{s=0}e^{-st/2}\mbox{deg}(n-3,s)~.
\eea

Now, sum up and again use Poisson summation to write in terms of deltas and take the integral, just as in~\cite{diaz-2009-42}. The indirect contributions to the trace are then collected in the following result
\be
\mbox{tr'}(\mathbb{S}^{-1}\partial_{\lambda}\mathbb{S}(\lambda))=-2l\sum^{\infty}_{N=1}\frac{e^{-\lambda lN}+e^{(\lambda-n)lN}}
{\mid1-e^{-lN+i\varphi N}\mid^2\times(1-e^{-lN})^{n-2}}.
\ee

\subsection{Full rotation matrix}

In case there is yet another block, one just has to replace the sum on the $(n-3)-$sphere in the very same way we did to add one block. That is, in the last factor of the trace formula of above (Eq.~\ref{eigen})

\bea\fl\nonumber\sum^{\infty}_{s=0}e^{-st/2}\mbox{deg}(n-3,s)\rightarrow
\sum^{\infty}_{m'=0}e^{-tm'}\;\left(1+\sum^{\infty}_{r'=1}e^{-r't/2}[e^{ir'\varphi' t/2l}+e^{-ir'\varphi' t/2l}]\right)\times\\
\times\sum^{\infty}_{s'=0}e^{-s't/2}\mbox{deg}(n-5,s')~.
\eea
Each sphere is being decomposed in smaller ones, just as nested dolls or \textit{matrioshkas}. Successive applications of this procedure finally lead to the expression for $K$ blocks:

\bea\fl
\mbox{tr'}(\mathbb{S}^{-1}\partial_{\lambda}\mathbb{S}(\lambda))=\\\nonumber
-2l\sum^{\infty}_{N=1}\frac{e^{-\lambda lN}+e^{(\lambda-n)lN}}
{\mid1-e^{-lN+i\varphi_1N}\mid^2...\mid1-e^{-lN+i\varphi_K N}\mid^2\,(1-e^{-lN})^{n-2K}}~,\eea
which is precisely the bulk result, as we will see,
\be
\frac{1}{2}\mbox{tr'}(\mathbb{S}^{-1}\partial_{\lambda}\mathbb{S}(\lambda))=
\frac{\mbox{Z'}_{\Gamma}}{\mbox{Z}_{\Gamma}}(\lambda)+
\frac{\mbox{Z'}_{\Gamma}}{\mbox{Z}_{\Gamma}}(n-\lambda)~,\ee
or, integrating,

\be
\mbox{det'}\,S(\lambda)=\left[\mbox{Z}_{\Gamma}(n-\lambda)/\mbox{Z}_{\Gamma}(\lambda)\right]^2~.
\ee

\section{Bulk}

For completeness, we examine the bulk side in order to parallel the boundary computation. The relative functional on this side of the correspondence can also be casted as a trace

\be\fl
\mbox{log}\,
\frac{\det_{-}(\Delta_X-\lambda(n-\lambda))}{\det_{+}(\Delta_X-\lambda(n-\lambda))}
=\mbox{tr}_-\mbox{\,log}\,(\Delta_X-\lambda(n-\lambda))-\mbox{tr}_+\mbox{\,log}\,
(\Delta_X-\lambda(n-\lambda))~.
\ee
We focus on the standard $+$branch and take the derivative with respect to $\lambda$ and compute in terms of the (truncated) heat kernel representation for the Green's function subtracting the direct contribution given by the Green's function for the original hyperbolic space. The trace here means taking the coincidence-point limit and integrating over the fundamental domain, so that
\be\fl
(2\lambda-n)\mbox{tr}{(G^+_X-G^+_H)}=(2\lambda-n)\int^{\infty}_0 ds\,\mbox{tr}\,K'_X(\sigma,s)\,e^{s\lambda(n-\lambda)}~.
\ee
The idea is to apply the method of images to compute the coincidence limit of the Green's function, that is, to sum up the contributions from image
points. We start with the heat kernel for the (positive) Laplacian $\mbox{exp}(-t\,\Delta_H)$ in hyperbolic space $\mathbb{H}^{N+1}$ can be compactly written in terms of Weyl's fractional derivative~\cite{AnkerAndOstellari}:

\be
\label{K}
K_{n+1}=\frac{e^{-tn^2/4}}{(2\pi)^{n/2}}\cdot\,_{_x}W^{\frac{n}{2}}_{_\infty}\left[K_1\right]~.
\ee
It is convenient to write it as a function of $x=\mbox{ch}\,\sigma$, where $\sigma$ is the geodesic distance, so that $x$ is essentially the chordal distance on the embedded hyperboloid (Eq.~\ref{Ddimensionalform}). The input is the heat kernel on the line
$K_1=\frac{1}{(4\pi t)^{1/2}}\,e^{-\sigma^2/4t}$.
Now, the geodesic distance between a point $(z,\vec{x})$ in the fundamental region and its $m-th$ image under the identification $(z_m,\vec{x}_m)=e^{ml}\,(z,\,\mathbb{A}^m\cdot\vec{x})$ satisfies
\be
\mbox{ch}\,\sigma(z,\vec{x}\mid z_m,\vec{x}_m)=\frac{z^2(1+e^{2ml})+\mid(\mathbb{I}-e^{ml}\mathbb{A}^m)\cdot\vec{x}\mid^2}{2z^2e^{ml}}~.
\ee

The fundamental region can be taken as $1\leq z \leq e^l$ with volume element $d^n\vec{x}\,dz/z^{n+1}$ and to take the trace of the heat kernel we need the volume integral of a function of the geodesic distance between image points. We proceed in two steps. First, change variables
\be
\vec{x}\rightarrow \vec{x}-\vec{x}_m\equiv\vec{y}_m=(\mathbb{I}-e^{ml}\mathbb{A}^m)\cdot\vec{x}
\ee
with Jacobian $\frac{\partial \vec{y}_m}{\partial \vec{x}}=\mathbb{I}-e^{ml}\mathbb{A}^m$. Second,
change variables again
\be
y_m \rightarrow \mbox{ch}\sigma_m=\frac{z^2(1+e^{2ml})+y_m^2}{2z^2e^{ml}}~.
\ee

The volume integral of a function of $u\equiv\mbox{ch}\sigma_m$ becomes then
\bea\fl
\int d \mbox{vol}_X\, \bullet&=&
\int^{e^l}_1\frac{dz}{z^{n+1}}\int^{\infty}_0 dy_m\;y_m^{n-1}\;\mbox{vol}(S^{n-1})\;\parallel\frac{\partial \vec{y}_m}{\partial \vec{x}}\parallel ^{-1} \bullet
\nonumber\\
\nonumber\\
&=&\int^{e^l}_1\frac{dz}{z^{n+1}}\int^{\infty}_0 dy_m\;y_m^{n-1}\;\mbox{vol}(S^{n-1})\;\parallel\frac{\partial \vec{y}_m}{\partial \vec{x}}\parallel ^{-1} \bullet
\\
\nonumber\\
\nonumber
&=&2^{n/2-1}\,e^{mln/2}\;\mbox{vol}(S^{n-1})\;|\mbox{det}(\mathbb{I}-e^{ml}\,\mathbb{A}^m)|^{-1}\times
\\\nonumber &\times&
\int^{e^l}_1\frac{dz}{z}\int^{\infty}_{ch\,ml} du\;(u-\mbox{ch}\,ml)^{n/2-1}\,\bullet~.
\eea

Here we notice that the last expression is nothing but a Weyl's fractional integral of order $n/2$, so that the volume integral of a function depending only on the geodesic distance between image points can be cast into the following convenient form
\be\fl
2^{n/2-1}\,l\,e^{mln/2}\;\mbox{vol}(S^{n-1})\;|\mbox{det}(\mathbb{I}-e^{ml}\,\mathbb{A}^m)|^{-1}
\,\Gamma(\frac{n}{2})\cdot\; _{_{ch\,ml}}W^{-\frac{n}{2}}_{_\infty}[\;\bullet\;]~.
\ee
Inserting the heat kernel for the quotient space, obtained by summing over image locations, the composition of fractional integral and derivative easily gives the indirect contributions to the trace of the heat kernel
\be\fl
\mbox{tr'}e^{-t\,\Delta_X}=l\sum_{m\in\mathbb{Z},m\neq0}\,e^{mln/2}\;|\mbox{det}(\mathbb{I}-e^{ml}\,\mathbb{A}^m)|^{-1}
\,\frac{1}{(4\pi t)^{1/2}}\,e^{-tn^2/4-(ml)^2/4t}~.
\ee

To get the trace of the Green's function, i.e. of the resolvent, it remains to take the proper-time integral. After straightforward manipulations and subtracting the analytically continued result from $\lambda_+=\lambda$ to $\lambda_-=n-\lambda$, we finally have
\be
\\
(2\lambda-n)\mbox{tr'}[G^+_X-G^-_X]=
-2l\sum_{m=1}^{\infty}\,\frac{e^{-\lambda ml}+e^{(\lambda-n) ml}}{|\mbox{det}(\mathbb{I}-e^{ml}\,\mathbb{A}^m)|^{-1}}~,
\ee
which is one of the many ways to express the log-derivative of the corresponding Selberg zeta function (cf.~\ref{AppendixZetaFunction})
\be
\\
(n-\lambda/2)\mbox{tr'}[G^+_X-G^-_X]=
\frac{\mbox{Z'}_{\Gamma}}{\mbox{Z}_{\Gamma}}(\lambda)+
\frac{\mbox{Z'}_{\Gamma}}{\mbox{Z}_{\Gamma}}(n-\lambda)~.
\ee
In terms of the determinants, the result is compactly expressed as

\begin{equation}
\frac{\det_{-}'(\Delta_X-\lambda(n-\lambda))}{\det_{+}'(\Delta_X-\lambda(n-\lambda))}
=\left[\mbox{Z}_{\Gamma}(n-\lambda)/\mbox{Z}_{\Gamma}(\lambda)\right]^2~.
\end{equation}

\subsection{Renormalized volume and Euler characteristic}

The prime on the determinants and traces above means that one has excluded the direct contribution, that is, the term containing the volume of the fundamental region in the bulk. This term requires renormalization, very much like in the exact hyperbolic case~\cite{Diaz:2007an},
\be\mbox{tr}{(G^+_H-G^-_H)}=coeff\times \{\textit{volume of fundamental region}\}~.
\ee
The finite coefficient in front is essentially the Plancherel measure for the hyperbolic space, but the infinite volume is that of the fundamental region.
Therefore, a renormalized volume for the fundamental is required.

There are some general results on this quantity, that apply in the conformally flat case that is considered here. When $n=odd$ and there are no torsion angles, i.e. $\mathbb{A}$ is trivial, the renormalized volume vanishes~\cite{diaz-2009-42}. This is supported
by the connection with the Euler characteristic $\chi$ of the conformally compactified bulk manifold $\bar{X}$, which vanishes in this case because the manifold is a solid torus~(cf.~\cite{Graham2007}). The inclusion of torsion angles does not change the topology and therefore the renormalized volume also vanishes.
When $n=even$, however, the renormalized volume is not conformal invariant and its value also differs for different renormalization schemes; we will assume that it vanishes, as in the non-spinning case when using dimensional regularization.

In all, sticking to dimensional regularization, the direct contribution vanishes and we can drop the primes in all previous formulas.

\section{Weierstrass regularization and quasi-normal frequencies}

Recently, a related holographic recipe has been given to compute the bulk determinant in terms of the spectrum of
quasi-normal modes~\cite{Denef:2009kn}. We will illustrate the connection to the holographic formula by examining the concrete example of $BTZ_3$.

Let us recall the expression for the holographic formula that we get in this case
\be
\frac{\det_{-}(\Delta_{BTZ}-\lambda(2-\lambda))}{\det_{+}(\Delta_{BTZ}-\lambda(2-\lambda))}
=\det\,S_T(\lambda)=\left[\frac{Z_{BTZ}(2-\lambda)}{Z_{BTZ}(\lambda)}\right]^2~.
\ee
The conformal boundary is a two-torus $T$ with a parallelogram as fundamental domain, described by a Teichm\"uller parameter, and
$Z_{BTZ}$ is the Selberg zeta function attached to the BTZ geometry by Perry and Williams~\cite{PerryandWillians},

\be
Z_{BTZ}(\lambda)=\prod_{k_1,k_2\geq0}\left[1-\alpha_1^{k_1}\alpha_2^{k_2}\,e^{-(k_1+k_2+\lambda)l}\right].
\ee
with $\alpha_1=e^{i\theta}$ and $\alpha_2=e^{-i\theta}$. The relation to the standard parametrization of the spinning BTZ$_3$ \cite{Banados:1993gq} is given by
\be
l= 2\pi r_+ \textrm{ and } \theta= 2\pi |r_{-}|.
\ee

It is direct to see that the set of zeros of $Z_{BTZ}(\lambda)$ is given by
\be\label{ZerosAndPoles}
\mathcal{R}=\left\{ \zeta_{k_1,k_2,m}=-(k_1+k_2)+i(k_1-k_2)\theta+2 i \pi \frac{m}{l} \right\}
\ee
with $k_{1,2}\in\mathbb{N}_0$ and $ m \in \mathbb{Z}$. However, the nontrivial observation\cite{PerryandWillians,Patterson00thedivisor} is that the set the poles of the scattering operator (Eq.(\ref{scattering})),
\be
\mathcal{R'}=\left\{s_{m,N,j}= -2j-|m| \pm i \frac{2\pi N - m \theta}{l}\right\},
\ee
with $j\in\mathbb{N}_0$ and $N,m\in \mathbb{Z}$, exactly matches the set in Eq.(\ref{ZerosAndPoles}). This can be casted as \textit{rudiments of holography}.

The Selberg zeta function $Z_{BTZ}$ has therefore a Hadamard product representation~\cite{PerryandWillians}, which can be thought of as a Weierstrass-regularized~\cite{Dowker:1995rh} version of the product of zeros,
\be
Z_{BTZ}(\lambda)=e^{Q(\lambda)}\,\prod_{\zeta\in \mathcal{R}}\,(1-\lambda/\zeta)\,e^{\lambda/\zeta+\frac{1}{2}(\lambda/\zeta)^2+\frac{1}{3}(\lambda/\zeta)^3}~,
\ee
where $Q$ is a polynomial of degree at most three with finite coefficients.
But due to the matching $\mathcal{R}=\mathcal{R}'$, one concludes then that the holographic formula produces a Weierstrass-regularized product of scattering resonances.

Now we come to the recipe in~\cite{Denef:2009kn} and to their main observation that, for the non-spinning BTZ, the set of scattering resonances can be rephrased as  Matsubara plus quasi-normal frequencies. To  see this, recall the spectrum of quasi-normal frequencies for a real scalar field in the non-spinning BTZ black hole
\be
\omega_{QN}=N-il(\lambda+2j)/2\pi~.
\ee
Notice then  that for the non-spinning BTZ
\be
\lambda-s_{m,N,j}=\lambda+2j+2\pi i N/l + |m| = |m|+2\pi i \omega_{QN}/l~,
\ee
so that the naive product of resonances can be rewritten in terms of the quasinormal frequencies as follows
\be
\prod_{\zeta\in \mathcal{R}}\,(\lambda-\zeta)\,=\,\prod_{s\in \mathcal{R}'}\,(\lambda-s)\,=\,\prod_{Matsubara, QN}\,(|m|+2\pi i \omega_{QN}/l)~.
\ee
The partial product over $m\in \mathbb{Z}$ (Matsubara frequencies) can be regularized using the gamma function, and the product can be casted in the form of product over quasinormal frequencies of gamma functions; however, the resulting expression is ill-defined because the infinite product is still divergent.
The Weierstrass regularization is a prescription to render these products finite and in the exact cases that were considered here, produces the Selberg zeta function.

An expression in terms of quasinormal frequencies is appealing because it can be extended beyond the exact cases considered here.
The proposal in~\cite{Denef:2009kn} is for the determinant in the bulk, whereas the holographic formula gives the answer with respect to a reference (the continuation to $n-\lambda$). The computation of the the separate determinants in the bulk requires the regularization/renormalization of the UV divergence in the bulk, which was bypassed by the holographic formula that matches IR in de bulk to UV on the boundary. Now, some caution must be taken in trying to separate the pieces of the holographic formula. The situation is in fact very reminiscent to the computation of the scalar field exchange Witten graph in the early days of AdS/CFT correspondence. The duality naturally gives an expression for the difference of exchange graphs involving bulk-to-bulk propagators for $\lambda$ and $n-\lambda$ which matches exactly the difference of the conformal partial waves of the dual operator with scaling dimension $\lambda$ and its conjugate with $n-\lambda$. Initially, assuming analyticity, each exchange amplitude and the corresponding conformal partial wave were claimed to be identical~\cite{Liu:1998ty}; however, it was later realized the presence of logarithmic terms that spoiled the identification~\cite{Liu:1998th}.

\section{Conclusion}

In this paper we have verified the holographic formula for the spinning BTZ geometry resulting from the quotient
with a full rotation matrix. The functional determinants are expressed in term of
the associated Selberg zeta function and the connection with quasi-normal frequencies has been elucidated.
It seems very likely that these functional determinants can be written in term of Selberg zeta function
for all cases where the spectrum of quasi-normal frequencies is explicitly known. The existence of quasi-normal modes
for spinor, vector and tensor excitations in the bulk strongly suggests that there must be suitable versions of the
holographic formula relating their one-loop bulk determinants to their boundary two-point functions.
The exact expressions in terms of the Selberg zeta function, more than just a mathematical curiosity, are suitable to
analytically explore the low- and high-temperature regimes, on one hand. On the other hand, much of the work done using WKB approximation to
the cases that deviate from the exact ones considered here, should correspond to the transition from  Selberg zeta function and trace formula
to the approximate ones given by a semiclassical zeta function and Gutzwiller trace formula (cf.~\cite{ChaosBook}).

\ack  This work was partially funded through Fondecyt-Chile 3090012.
\appendix

\section{Constant curvature spaces}\label{Coordinates}

In general a constant curvature space can be written in terms of

\begin{equation}\label{GeneralConstCurvatureSolution}
ds^{2} = B(\rho)^2 (\hat{e}^{i}\hat{e}^{j}\eta _{ij})+C(\rho)^{2}d\rho^{2}+D(\rho)^{2}(\tilde{e}^{m}\tilde{e}^{n}\eta _{mn}),
\end{equation}
or in terms of the vielbein
\begin{eqnarray*}
e^{i} &=&B(\rho)\hat{e}^{i} \\
e^{r} &=&C(\rho)d\rho \\
e^{n} &=&D(\rho)\tilde{e}^{n}.
\end{eqnarray*}
Since this is a torsion-free solution, $T^{a}=0$, therefore the connection is given by
\begin{eqnarray*}
w^{ij} &=&\hat{w}^{ij} \\
w^{ir} &=&\frac{B(\rho)^{\prime }}{C(\rho)B(\rho)}e^{i} \\
w^{mr} &=&\frac{D(\rho)^{\prime }}{C(\rho)D(\rho)}e^{m} \\
w^{mn} &=&\tilde{w}^{mn},
\end{eqnarray*}
and so the curvatures are completely determined by the intrinsic curvatures as
\begin{eqnarray*}
R^{ij} &=&\hat{R}^{ij}-\left( \frac{\ln (B(\rho))^{\prime }}{C(\rho)}\right)
^{2}e^{i}\wedge e^{j} \\
R^{ir} &=&-\frac{1}{B(r)C(r)}\left( \frac{B(\rho)^{\prime }}{C(\rho)}\right)
^{\prime }e^{i}\wedge e^{r} \\
R^{im} &=&-\frac{\ln (B(r))^{\prime }\ln (D(\rho))^{\prime }}{C(\rho)^{2}}%
e^{i}\wedge e^{m} \\
R^{rm} &=&-\frac{1}{A(r)C(r)}\left( \frac{D(\rho)^{\prime }}{C(\rho)}\right)
^{\prime }e^{r}\wedge e^{m} \\
R^{mn} &=&\tilde{R}^{mn}-\left( \frac{\ln (D(\rho))^{\prime }}{C(\rho)}\right)
^{2}e^{m}\wedge e^{n}.
\end{eqnarray*}

This solution has still the invariance of the definition of the $\rho$ coordinate. By fixing $C(\rho)= l (2\rho)^{-1}$, as in the usual Fefferman-Graham coordinates, the constant curvature solution, \textit{i.e.} $\bar{R}^{ab}=R^{ab}+ l^{-2} e^{a} e^{b}=0$, is given by
\begin{eqnarray*}
  B(\rho) &=&  \frac{l \sqrt{2}}{4\sqrt{\rho\beta(c_1-c_2)}} (\rho(c_1-c_2)+2)\\
  D(\rho) &=&  \frac{l \sqrt{2}}{4\sqrt{\rho\alpha(c_2-c_1)}} (\rho(c_2-c_1)+2)\\
\end{eqnarray*}
where the intrinsic curvatures are given by,
\begin{equation*}
  \hat{R}^{ij} = -\frac{1}{\beta}\hat{e}^{i} \hat{e}^{j} \textrm{ and }
  \tilde{R}^{mn} = -\frac{1}{\alpha} \tilde{e}^{m} \tilde{e}^{n}
\end{equation*}
with $\alpha=-\beta=\pm 1$ and $c_{1}$ and $c_{2}$ are constant to be determined.

\section{From the quadratic line element to Poincare coordinates}
In order to write explicitly the identifications defined by Eq.(\ref{Identification}) is useful to write the quadratic form
\begin{equation}\label{Ddimensionalform}
-p^2 + w^2 + (u^1)^2+\ldots +(u^{d-1})^2 = -1.
\end{equation}
The Poincare coordinate set for $H_{d}$ arises from
\begin{eqnarray*}
  p &=& \frac{1}{2z}\left(z^2+(y^1)^2+\ldots+ (y^d) + 1 \right) \\
  w &=&  \frac{1}{2z}\left(z^2+(y^1)^2+\ldots+ (y^d) -1 \right) \\
  y^{i} &=& \frac{u^i}{2z}
\end{eqnarray*}
with $i=1\ldots d-1$. The Poincare half plane is therefore described by
\begin{equation}\label{PoincareCoordinatesd}
ds^2 = \frac{1}{z^2} (dz^2 + \delta_{ij} dy^i dy^j)
\end{equation}
reproducing Eq.(\ref{Poincare}).

In this coordinates the boost in the plane $w,p$, generated by the Killing vector
\begin{equation}\label{SimplestKilling}
\xi = w\frac{\partial }{\partial p} + p\frac{\partial }{\partial w},
\end{equation}
is given by
\[
\xi = z\frac{\partial }{\partial z} + y^i \frac{\partial }{\partial y^i}.
\]

Obviously the generator $g = e^{l \xi}$ represents a dilatation in Eq.(\ref{PoincareCoordinatesd}).

On the other hand on can notice that the set of all rotation that commute among themselves and with $\xi$ is given, up to some global rotation, by rotations in each of the planes $(u^{k_{1}},u^{k_{2}})=\{(u^1,u^2),(u^3,u^4),\ldots,(u^{j-1},u^{j})\}$, with $j$ the integer part of $(d-1)/2$. This rotations are generated by
\begin{equation}
\zeta_{k} = \left(u^{k_{1}}\frac{\partial}{\partial u^{k_{2}}} - u^{k_{2}}\frac{\partial}{\partial u^{k_{1}}}\right),
\end{equation}
which in Poincare coordinate can written as
\[
\zeta_{k} = \left(y^{k_{1}}\frac{\partial}{\partial y^{k_{2}}} - y^{k_{2}}\frac{\partial}{\partial y^{k_{1}}}\right).
\]

\section{Selberg zeta function}\label{AppendixZetaFunction}

The Selberg zeta function associated to the quotient geometries we consider was first introduced by Patterson~\cite{PattersonKleian89} in the form of Euler product. In terms of the length $l$ of the (primitive) closed geodesic and the eigenvalues $\{\alpha_1,...,\alpha_n\}$ of the rotation matrix $\mathbb{A}$, it is given by
\be
Z_{\Gamma}(\lambda)=\prod_{k_1,...,k_n\geq0}\left[1-\alpha_1^{k_1}...\alpha_n^{k_n}\,e^{-(k_1+...+k_n+\lambda)l}\right]~.
\ee
This Selberg zeta function has also an interpretation as dynamical zeta function for geodesic flow on $X=\Gamma\backslash\mathbb{H}^{n+1}$~\cite{PerryandWillians}. The matrix $\mathbb{A}$ describes the rotation of nearby closed geodesics under the Poincare once-return map $\mathcal{P}_{\gamma}$. Elementary manipulations lead to
\be
\log Z_{\Gamma}(\lambda)=-\sum_{m\geq1}\frac{1}{m}\,\frac{e^{-m\lambda l}}{\det(\mathbb{I}-e^{-ml}\mathbb{A}^m)}~,
\ee
with
\be\fl
\det(\mathbb{I}-e^{-ml}\mathbb{A}^m)=\prod_{j=1}^{n}(1-\alpha^m_j\,e^{-ml})=e^{-mln/2}\,|\det(\mathbb{I}-\mathcal{P}^m_{\gamma})|^{1/2}~.
\ee

\vspace{0.5cm}

\section*{References}


\providecommand{\href}[2]{#2}\begingroup\raggedright\endgroup

\end{document}